# Designing contrasts for rapid, simultaneous parameter quantification and flow visualization with quantitative transient-state imaging

**Pedro A. Gómez [1], Miguel Molina-Romero [1], Guido Buonincontri [3],**

**Marion I. Menzel [2], Bjoern H. Menze [1]**

[1] Technical University of Munich, Munich, Germany

[2] GE Healthcare, Munich, Germany

[3] Imago7 Foundation, Pisa, Italy

Corresponding author:

Dr. Pedro A. Gómez

Munich School of Bioengineering

Technical University of Munich

Boltzmannstr. 11, D-85748 Garching, Germany

E. pedro.gomez@tum.de

# Abstract

Magnetic resonance imaging (MRI) is a remarkably powerful diagnostic technique: it generates wide-ranging information for the non-invasive study of tissue anatomy and physiology. Complementary data is normally obtained in separate measurements, either as contrast-weighted images, which are fast and simple to acquire, or as quantitative parametric maps, which offer an absolute quantification of underlying biophysical effects, such as relaxation times or flow. Here, we demonstrate how to acquire and reconstruct data in a transient-state with a dual purpose: 1 – to generate contrast-weighted images that can be adjusted to emphasise clinically relevant image biomarkers; exemplified with signal modulation according to flow to obtain angiography information, and 2 – to simultaneously infer multiple quantitative parameters with a single, highly accelerated acquisition. This is a achieved by introducing three novel elements: a model that accounts for flowing blood, a method for sequence design that incorporates both parameter encoding and signal contrast, and the reconstruction of temporally resolved contrast-weighted images. From these images we simultaneously obtain angiography projections and multiple quantitative maps. By doing so, we increase the amount of clinically relevant data without adding measurement time, creating new dimensions for biomarker exploration and adding value to MR examinations for patients and clinicians alike.

# Introduction

**MR contrasts versus parameter quantification.** Since Lauterbur's breakthrough idea[1], magnetic resonance imaging (MRI) has enjoyed decades of incremental improvements, evolving into an unparalleled imaging modality with the ability of providing detailed information on a tissue's structure and function. Modern MR scanners use sophisticated combinations of radiofrequency pulses and magnetic gradients to probe the complex dynamics of spins inside the human body. In its simplest form, scanners produce images 'weighted' by the longitudinal relaxation time (T1), the transverse relaxation time (T2), and the proton density (PD). However, contrast-weighted images are subjective: the expertise of radiologists plays a key role in their evaluation for disease diagnosis and monitoring. To this end, advanced mapping techniques have been developed – all based on series of weighted acquisitions – that offer means for an absolute quantification of biophysical phenomena, such as tissue relaxation time or displacement velocity, and that help in increasing accuracy and reproducibility of diagnostic information[2]. Notwithstanding, it is contrast, not quantification, which has allowed MRI to thrive as one of the most important diagnostic modalities.

**Advantages of qualitative MR contrasts.** The fundamental reason for the success of contrast – i.e. relative local differences in qualitative images – over quantification can be attributed to the simpler process of signal encoding: whereas contrast-weighted MRI merely *accounts* for physical effects interacting with the acquired signal to produce clinically relevant images, quantitative MRI aims at *encoding* for these. This adds complexity to the acquisition: First, additional samples need to be acquired along multiple encoding dimensions, consequently increasing scan times, sometimes beyond clinical acceptance. Second, most quantitative approaches probe only a single or a few parameters at a time, requiring lengthy serial acquisitions if multiparametric information is required. Finally, quantitative parameters are estimated by enforcing consistency with a biophysical model, which, by definition, is subject to limitations.

As an example, MR angiographic scans are fast and simple to acquire, and even though they do not quantify the velocity or direction of flowing blood in an absolute fashion, they represent an indispensable biomarker in the diagnosis of strokes[3], stenosis[4], and other vascular diseases. 4D flow MRI[5], the quantitative counterpart of MR angiography, is significantly more involved: it requires multiple directional velocity encoding gradients to capture time-varying velocity vectors and a corresponding high-dimensional, spatiotemporal biophysical model to accurately measure and visualize complex blood flow patterns. While promising, the cost in terms of increased acquisition time and the uncertainty of the resulting flow quantification still prevents the widespread clinical adoption of 4D flow techniques[6]. Thus, with flow – as with other quantitative techniques – the current provider of clinical value remains contrast rather than quantification.

**The advantage and the shortcomings of novel quantitative sequences.** The recent advent of fast multiparametric mapping methods offers new directions for challenging the longstanding dominance of contrast-weighted imaging biomarkers. Novel techniques such as MR fingerprinting (MRF)[7], synthetic MRI[8], and others[9–13] promise to ease the clinical implementation of quantitative imaging techniques by developing sequences to quickly deliver multiparametric information with a single scan. Recent advances offer high-resolution maps with ground-breaking speed[14], creating new clinical opportunities for accelerated, quantitative mapping. These methods have already been extended to consider the effects of blood flow by deriving vascular parameters after contrast injection[15,16], computing velocity scalars[17] or vectors in phantom studies[18]. However, quantitative techniques share a common drawback: they are designed to exclusively provide the information derived from the biophysical model, and since quantitative information obtained from simplified modelling is subject to limitations, could disregard valuable diagnostic information.

**Our contribution: designing contrasts with a quantitative sequence.** Here, instead of limiting the acquisition to obtain exclusively quantitative parameters, we propose to reconstruct spatially resolved temporal signals that accommodate additional sources of information, opening the door to new possibilities for data analysis and biomarker exploration. Our solution is inspired by the promising advances shown by MRF, but removes its two signature features – namely a pseudorandom acquisition and the matching of signals to a precomputed dictionary – and replaces them with a sequence design and parameter inference framework to achieve a dual purpose: 1 – create contrasts for clinically relevant image biomarkers, and 2 – infer multiple quantitative maps – both simultaneously with a single scan. We term our method 'quantitative transient-state imaging' and achieve our dual purpose by combining three novel elements:

1. First, our solution builds a ***transient-state model*** based on the latest multiparametric advances for quantitative encoding, while also ***considering motion phenomena*** that perturb transient signals during acquisition time.

2. Second, it offers a ***design framework for extracting diagnostically relevant contrasts*** from these physiological effects during parameter encoding.

3. Third, it relies on a ***4D reconstruction to create hundreds of contrast-weighted images*** from which relevant image biomarkers can be obtained ***alongside the quantitative parameters*** of the Bloch equations.

Moreover, the parameters are inferred probabilistically from the model in the transient-state, resulting in additional estimates of parametric uncertainty. In addition to these inferred parametric maps, the reconstructed images provide complementary physical information which can be tuned to visualize – and emphasise – specific physiological features like blood volumes or tissue types. In this way, we maintain the advantages of clinically successful contrast-weighted 'qualitative' imaging using a sequence that is intrinsically quantitative in nature.

We illustrate QTI with the aforementioned example of flow, where we modulate signals in the transient-state to encode for relaxation times while simultaneously accounting for the effects of flowing blood. In this manner, we can obtain – on the one hand – multiple contrast-weighted images, including specifically designed contrasts that carry the information of MR angiography images; and – on the other – fast, high-resolution, and accurate measurements of the encoded parameters. By doing so, we add vasculature biomarkers without increasing scan times and offer novel prospects for precision medicine applications, such as designing disease- and biomarker-specific imaging sequences.

## Materials & Methods

**Formalising spin dynamics in the transient-state.** A strategy to encode for multiple parameters, demonstrated in MRF[7], is to avoid the steady-state by continuous variations of the acquisition variables. In this transient-state, the magnetization vector $\mathbf{M}_t = [M_x(t)\ M_y(t)\ M_z(t)]^\top$ evolves dynamically in time and, following the Extended Phase Graph (EPG)[19,20] formalism, a model of magnetization over time $\mathbf{M}_t(\eta;\theta)$ can be defined in a recursive manner:

$$\mathbf{M}_t(\eta;\theta) = \mathbf{M}_{t-1} \cdot g(\eta;\theta). \qquad [1]$$

Per EPG, the magnetization is interpreted as Fourier configuration states, where the magnetization $\mathbf{M}_t(\eta;\theta)$ at a given time $t$ is determined by the magnetization at time $t-1$ modulated by the operator $g(\cdot)$, which results from the concatenated application of physical operations affecting the magnetization vector, such as radiofrequency excitation, relaxation or gradient dephasing (a detailed EPG review can be found in Weigel *et al.*[19]). The final operator, $g(\eta;\theta)$, depends on two variable sets: $\eta$, the design variables of a potential acquisition scheme (e.g. flip angle or repetition time); and $\theta = \{T1, T2\}$, the spatially varying biophysical parameters we wish to infer.

We propose to extend this signal model to account for flow by assuming that spins will washout from the imaging plane[21] after being tagged with a radiofrequency pulse $i$:

$$\mathbf{M}_t(\eta; \theta) = \sum_{i}^{T} \mathbf{M}_{t-1} \cdot g(\eta; \theta) \cdot h_i(t), \qquad [2]$$

where $h_i(t)$ gives the fraction of spins labelled with pulse $i$ remaining in the imaging plane after time $t$. The function $h_i(t)$ is applied as an element-wise operation for every pulse $i$ over the entire dynamic time-course of the magnetization (Fig. 1). When there is no flow, i.e. $h_i(t) = 1$, spins become saturated from the continuous application of radiofrequency pulses and follow the dynamics dictated by the choice of acquisition parameters $\eta$. In the presence of flow, unsaturated or 'fresh' spins in every repetition wash into the imaging slice, perturbing the signal throughout the acquisition. It is important to note that $h_i(t)$ accounts exclusively for flow perpendicular to the imaging slice and is agnostic to the direction of flow and can therefore be used to capture both venous and arterial flow. We take advantage of this model to design a sequence to optimally encode for $\theta$ while allowing signals to evolve towards clinically relevant tissue contrasts.

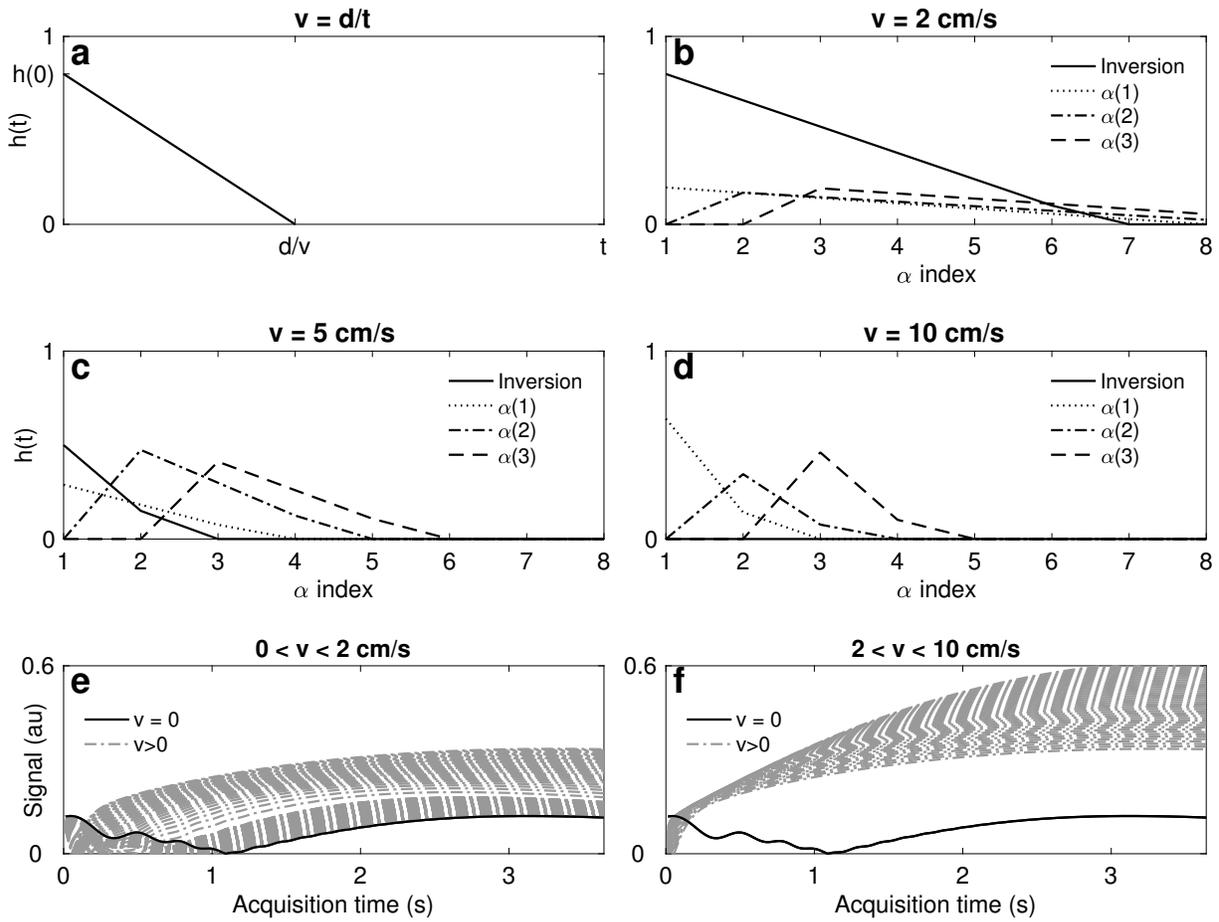

*Figure 1. Velocity effects with multiple pulses in the transient-state. **a**, Fraction of spins $h(t)$ remaining in the imaging slice after being tagged by a radiofrequency pulse at $h(0)$, where this washout function is given by the ratio of slice thickness $d$ to the velocity of flow $v$. **b-d**, The total number of repetitions that spins tagged by a certain pulse $\alpha$ remain in the slice is proportional to their velocity, where slow flowing spins will remain in the slice longer than faster spins. Each $\alpha$ pulse has its own washout function, where the initial fraction of spins is determined by the on the amount of unsaturated spins that have flown into the slice since the previous pulse. **e-f**, Stationary blood (solid black line) acts as a lower bound for blood flow contrast. Even for slow flowing spins, after the inversion non-stationary blood always have larger signal intensities than stationary blood spins.*

**Bayesian experimental design for contrast maximisation with optimal encoding.** Even though it is possible to use random and arbitrary patterns of $\eta$ to encode for $\theta$, recent optimisation work has demonstrated that experimental efficiency can be significantly increased by designing the acquisition scheme to satisfy specific criteria[22–25]. Here, we rely on Bayesian decision theory to guide our experimental design[26,27]; where we aim at finding an acquisition scheme that maximises the expected information gain for a prior probability distribution of parameters.

Let the complex signal over time be determined from the magnetization vector as $f_t(\eta; \theta) = M_x(t) + iM_y(t)$. The goal of Bayesian experimental design is thus to identify the optimal set of design variables $\eta$ that maximises the encoding of the biophysical parameters of interest $\theta$ for a given prior distribution of these parameters. This is achieved by selecting the design variables that maximise the information gain of the parameters. Assuming a Gaussian noise model, the information gain for a particular $\theta$ set can be represented as a utility functional equivalent to the determinant of the Fisher information matrix:

$$u(\eta; \theta) = \det \sum_t \begin{pmatrix} \frac{\partial^2 f_t}{\partial \theta_1^2} & \cdots & \frac{\partial^2 f_t}{\partial \theta_1 \theta_N} \\ \vdots & \ddots & \vdots \\ \frac{\partial^2 f_t}{\partial \theta_N \theta_1} & \cdots & \frac{\partial^2 f_t}{\partial \theta_N^2} \end{pmatrix}. \quad [3]$$

Note that a Gaussian noise model is a simplifying assumption, as $f_t(\eta; \theta)$ is acquired as a complex vector and therefore its magnitude follows a Rician noise distribution. Nonetheless, this Bayesian framework resulted advantageous despite the simplifying assumption, as it allows for the definition of the utility functional, and in a Bayesian manner, the overall utility can be determined by marginalisation the functional over a prior $\pi(\theta)$:

$$U(\eta) = \int_\theta \log u(\eta; \theta) \pi(\theta) d\theta. \quad [4]$$

Finally, the contrast created by the mean values of the different tissue classes in the prior can be incorporated into a final cost function:

$$C(\eta) = -\left((1 - \sum \mu_c)U(\eta) + \sum_{c=1}^{C} \mu_c \max_{f_t} |f_t(\eta; \theta_c) - f_t(\eta; \theta_d)| \ \forall \ d \neq c\right). \quad [5]$$

In Eq. 3, the cost function $C(\eta)$ is minimised when the overall utility $U(\eta)$ is maximised and the contrast for each individual tissue class $c$ is maximised with respect to all other tissue classes, where $\mu_c$ controls the contribution of each tissue class and the overall weighting of utility maximisation versus contrast maximisation. Since Eq. 3 is non-convex, global optimisation techniques, such as genetic algorithms, are required. If the dimensionality of the design space is small enough, even exhaustive searches can be performed, as the optimal design variables $\eta$ need to be determined only once and the optimisation can be performed offline.

We assumed a broad Gaussian distribution of four tissue classes using the following mean T1/T2 values: 1,450/85 ms for grey matter (GM), 900/60 ms for white matter (WM), 3600/1750 for cerebrospinal fluid (CSF), and 1740/275 for stationary blood (Fig. 2a). We started by inverting the magnetization with an inversion pulse and found the optimal flip angle design $\eta$ by setting two control points (minimum and maximum flip angle) and solving for Eq. 5 using a normalized $\mu_c =$ [0.05 0.05 0.1 0.3] for GM, WM, CSF and SB respectively. To derive a computationally tractable solution in parameter space, we followed the approach described in Owen *et al.*[27], while we additionally visualize the optimization landscape in Figs. 2b-g. The mean tissue values were only considered for the stationary case ($h_i(t) = 1$), as in the domain of linear flip angle variations, the contrast for stationary blood after the inversion acts as a lower bound for flowing blood after inversion (Fig. 1e-f). Certainly, the same design framework can be used to find other higher-order flip angle patterns with more control points (Supplementary Fig. 1).

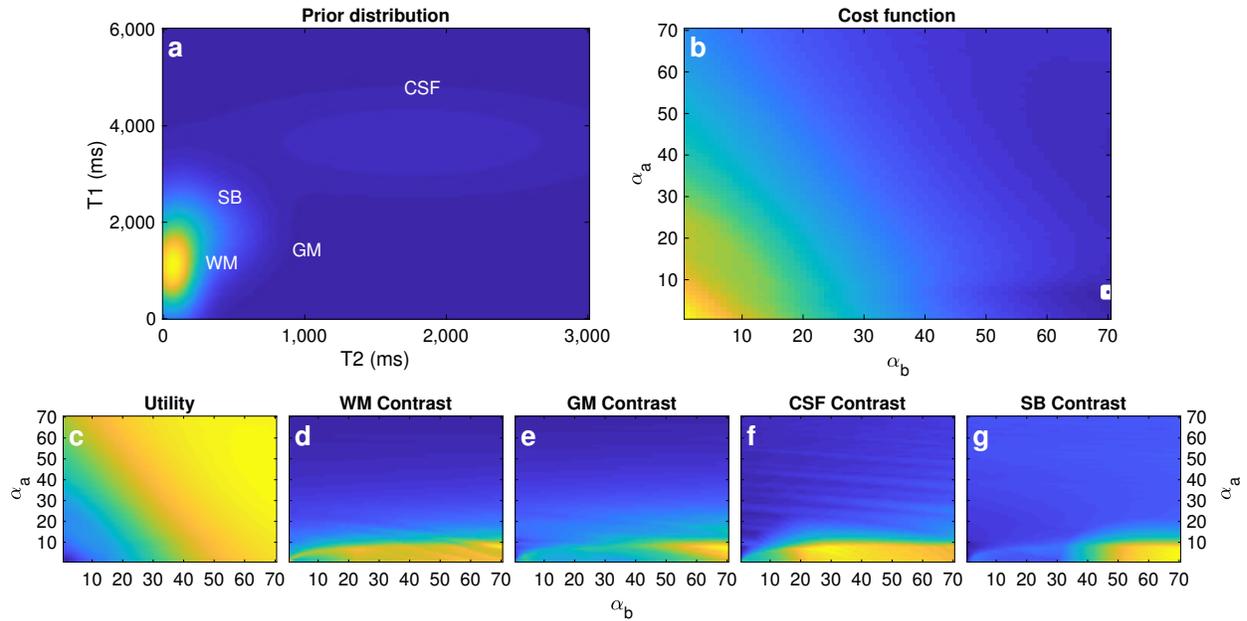

*Figure 2. Constrained Bayesian experimental design. a,* T1 and T2 space with Gaussian priors of four tissue classes: grey matter (GM), white matter (WM), cerebrospinal fluid (CSF), and stationary blood (BS) *b*, Visualization of optimization landscape with respect to the initial ($\alpha_a$) and final ($\alpha_b$) flip angle. The cost function reaches a minimum when $\alpha_a/\alpha_b$ is 7/70 degrees (white square). *c-g,* Contribution of the individual terms to the final cost function. Whereas the utility is larger for larger initial and final flip angles, contrast is maximised when the initial flip angle is $\alpha_a <$ 10 degrees. The weighting of each individual cost function term is application specific and can result in different designs for distinct use cases.

The resulting design in the transient-state consists of an inversion pulse followed by a variable flip angle ramp (Fig. 3a) with 260 repetitions, constant repetition and echo times at TE/TR = 2/14 ms, 2 mm thick slices, and an unbalanced gradient moment in each repetition (Supplementary Fig. 2). This encoding strategy, while seemingly simple, is highly efficient: the first portion of the sequence encodes mostly for T1 as the magnetisation recovers from the inversion with T1 relaxation, and the last segment encodes mostly for T2, as higher flip angles allow for the formation of stimulated echoes. Hence, tissues with longer T1 will have an inversion later in time (Fig. 3c) and tissues with shorter T2 will cause smaller stimulated echoes in subsequent excitations (Fig. 3d). Moreover, contrast maximisation for SB towards the end of the acquisition combined with the thin imaging slices amplify flow effects, from which we recover angiography information.

Finally, the proposed design results in smooth signal variations with respect to $\theta$, leading to a certain level of correlation between an ensemble of multiple transient-state signals (Fig. 3f). That is, similar tissue types will have similar signal evolutions, while different tissues will distinguish form each other throughout the course of the acquisition. We explicitly rely on the correlation between signals to reconstruct the spatiotemporal image space via lower dimensional signal constraints.

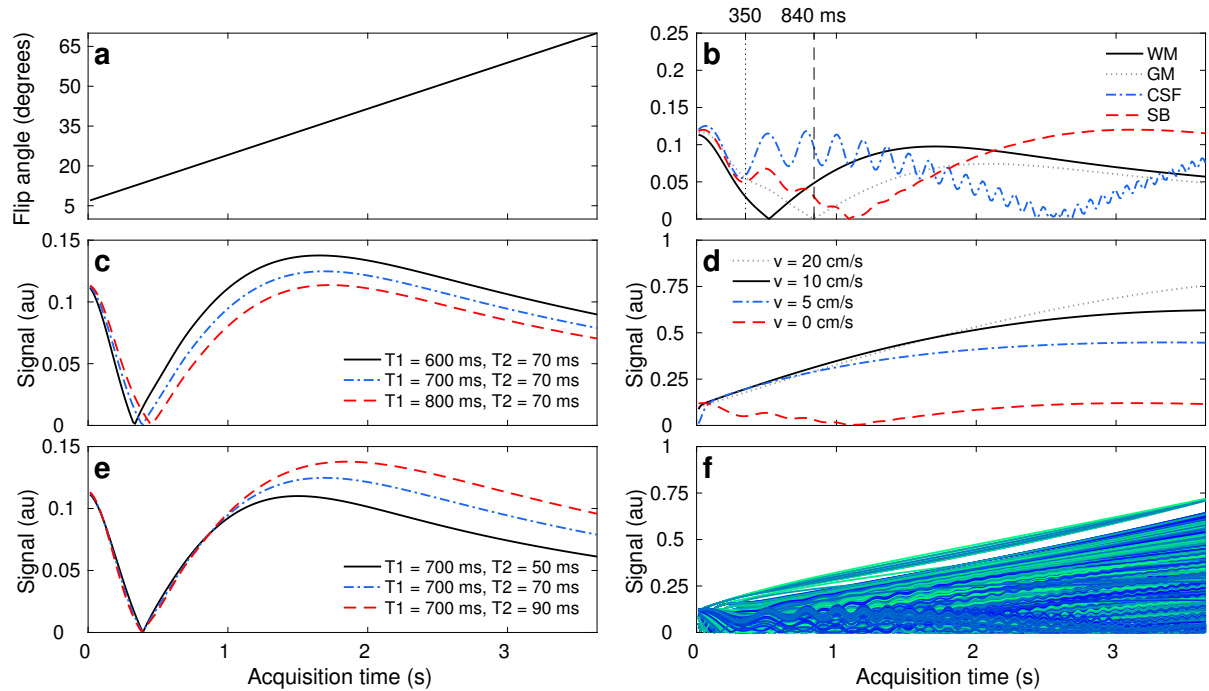

*Figure 3: Parameter encoding scheme and signal dynamics in the presence of flow. a, Flip angle ramp corresponding to the encoding scheme. b, Signal evolution from the mean value of the four classes: grey matter (GM), white matter (WM), cerebrospinal fluid (CSF), and stationary blood (SB). The sequence has been designed to have distinct signal evolutions for each class; where SB contrast discriminates well from the others towards the end of the sequence. Our proposed reconstruction framework also enables the visualisation of tissue dynamics through time, where e.g. WM/GM tissue contrasts at 350 and 840 ms are inverted as they recover from the inversion pulse with different T1 relaxation times. c,e, Signal evolutions for varying T1/T2 values, where (c) longer T1s (dashed, red line) experience their inversion later in time and (e) shorter T2s (continuous, black line) produce more signal decay throughout the course of the experiment. d, Signal hyperintensities due to fast flowing spins with 2 mm slice thickness. The thinner the imaging slice or the faster the flow, the more pronounced flow effects become. f, Ensemble of signals simulated from a prior parameter distribution of stationary and non-stationary tissues. As signals exhibit correlation in space and time, it is possible to use them to create a lower dimensional subspace for image reconstruction. The proposed design scheme satisfies maximal contrast conditions and optimal parameter encoding in 3.66 seconds per imaging slice.*

**4D spatiotemporal image reconstruction.** As the signal evolves in the transient-state, we acquire undersampled spatial data with a spiral arm in every repetition (Supplementary Fig. 2), since acquiring the full temporal signal for every 3D voxel in the image would increase the scan time to several hours[28]. We then rely on the theory of compressed sensing to recover full images from the measurements by incorporating prior knowledge into the reconstruction formalism. By recovering images, we take a similar line to other techniques which have proved successful for different kinds of dynamic MRI data[29–32], including MRF[22,33–35]; whereas we rely on local correlations from the entire 4D spatiotemporal neighbourhood to obtain a series of reconstructed images for every point in time.

Specifically, we formulated the image reconstruction problem to account for both the temporal spin dynamics and the Fourier relationship of the spatial signal, where the acquired image $x_t$ at every point in time is related to the acquired measurements $y_t$ by an encoding operator $y_t = E_t x_t$; which in turn consists of three terms: $E_t = U_t F S$. $U_t$ represents the spatial acquisition trajectory – a spiral waveform obtained with time-optimal gradient design[36], $F$ is the non-uniform fast Fourier transform[37], and $S$ are the coil sensitivities, obtained via adaptive coil combination of the acquired data[38]. By additionally incorporating a temporal subspace projection[22,30,34] operation into the encoding operator, we used the alternating direction method of multipliers[30,34] to reconstruct ten subspace images, regularized via local low rank thresholding on spatiotemporal images patches[30,35] of dimension 8 x 8 x 8 x 260. We then projected the subspace images back to the temporal space to obtain the full spatiotemporal image space.

**Angiographic projections and probabilistic parameter inference.** Obtaining spatially resolved MR signals allows to us to perform different forms of analysis on the reconstructed data. On the one hand, a qualitative combination of different time frames enables the creation of angiography images. This combination is achieved by summing the magnitude images of chosen frames and subsequently performing a maximum intensity projection. This qualitative assessment allows for a visualization of imaging biomarkers – such as those given by angiography – that may not be readily available in the quantitative parametric maps.

On the other hand, a quantitative analysis of the data allows us to simultaneously infer PD, T1, and T2. An important distinction to previous work in parametric mapping is that we do not perform inference by matching to a dictionary of precomputed signals, but use Bayesian inference for uncertainty quantification and propagation. We used $\Pi 4U^{39}$, a high-performance computing tool that relies on Transitional Markov Chain Monte Carlo sampling, to compute the posterior probability density function $p(\theta|x_t, f_t)$ of the parameters given the reconstructed data $x_t$ and the signal model $f_t$:

$$p(\theta|x_t, f_t) = \frac{p(x_t|\theta, f_t)\pi(\theta)}{p(x_t|f_t)}. \qquad [6]$$

In Eq. 6, $p(x_t|\theta, f_t)$ is the likelihood of observing the data from the model, $\pi(\theta)$ is the prior, and $p(x_t|f_t)$ is the evidence of the model. For Bayesian inference, unlike Bayesian experimental design, we assume a uniform prior in parameter space to avoid biasing parameter quantification. From the probability density function, we can obtain the maximum likelihood of each of the parameters in the model, the expected value, and the corresponding uncertainty. The introduction of uncertainty quantification with the probabilistic inversion of the transient-sate model provides complementary voxel-wise information on the correspondence between the reconstructed signals and the estimated maps. Uncertainty quantification also gives an indication of consistency of the measurements regarding the model, resulting in additional feedback on the encoding capabilities of the sequence with respect to each individual parameter.

**Data acquisition and analysis.** Important benchmarks for parametric mapping techniques are the accuracy and efficiency of the obtained measurements. Thereafter, we obtained reference measurements from agar phantoms using gold-standard T1 and T2 techniques and compared them against QTI and the first MRF implementation with unbalanced gradients[40] which been used in multiple subsequent works[28,34,41–43], as it is robust to artifacts and has been extensively validated for reproducibility[44]. We also shortened the MRF acquisition from 1000 repetitions to 280, resulting in the same scan time as QTI and refer to this shortened version as MRF*.

We scanned the Eurospin T05 phantom, comprising of different vials with characteristic T1 and T2 values to perform the accuracy and efficiency analysis. The gold standard was obtained with inversion recovery and multi-echo spin echo acquisitions for T1/T2 quantification. The measurements were done with 22.5 x 22.5 $cm^2$ field of view, 128 x 128 matrix size, 5 mm thickness, TR = 12,000 ms, TI = {2,700, 680, 300, 1,950, 1,320, 3,745, 815, 3,490, 2,000, 1,700, 950, 550, 3,600, 2,600, 3,100, 3,900, 3,400, 3,230, 2,350, 1,800, 50, 2,850, 170, 430, 2,210, 2,980, 1,580, 1,000, 2,470, 1,200, 1,500} ms, and TE = {20, 300, 100, 600, 75, 200, 35, 400, 150, 50} ms. Both the TI and TE were acquired in this random order to minimise magnetic drift and temperature effects. We obtained 10 independent measurements for QTI, MRF, and MRF*, the truncated version of MRF to 3.66 seconds (280 repetitions). For each tube, we selected a central region and obtained a total of 5,000 samples over the 10 measurements to perform statistical analysis. Both the concordance correlation coefficient and efficiency were estimated per the original MRF publications[7,40]. The concordance correlation coefficient estimates the average similarity of the quantification for all tubes with respect to the reference. The efficiency is defined as the precision (mean / standard deviation) divided by the square root of the acquisition time. The time of each sequence considered for the efficiency analysis was 3.66 s for QTI and MRF* (3.64 s readout plus 0.02 s inversion pulse) and 13.13 s for MRF (13.11 s readout plus 0.02 s inversion pulse).

We also scanned 112, 2 mm slices of two healthy volunteers (male, 28 and 26 years) with the proposed method (acquisition time = 6:48) and with a 3D axial time of flight angiography sequence with TE/TR = 2.1/24, 22 x 22 cm$^2$ field of view, 256 x 256 matrix, acceleration factor 2 in phase, and acquisition time = 10:48. The in vivo scans were conducted complying with the German Act on Medical Devices without requiring IRB approval, as our methodological work does not amount to a clinical investigation and the volunteers (experts in MRI and researchers in our team) where fully aware of potential risks. Written informed consent was obtained for each volunteer and all experiments were performed on a machine that has met all safety and functionality requirements for its intended purpose. For all scans, data was acquired with a single arm of a variable density spiral waveform within each TR. Each waveform required 18 interleaves to sample the centre of k-space and 89 to sample a full 22.5 x 22.5 field of view, resulting in 1.2 mm$^2$ in-plane resolution. To increase sampling incoherence[45,46], the waveforms were rotated with the golden angle from one repetition to the next. All experiments were performed in a single session on a 3T 750w scanner (GE Healthcare, Milwaukee, WI), with a 12-channel head receiver-only coil.

**Code and data availability.** The code used to simulate the transient-state signals, reconstruct multiple images, and estimate parametric maps is available here: https://bitbucket.org/pgomezd/qti. The source code for $\Pi 4U$ can be downloaded from the authors' website: http://www.cse-lab.ethz.ch/software/Pi4U. A simulated phantom is available in the code repository. The phantom can be used to test the algorithms presented in this study. All additional data supporting the findings of this study are available within the paper and its Supplementary Information.

# Results

**Phantom study.** Figures 4a and 4b show that both QTI and MRF are in good agreement to the reference, achieving T1/T2 correlation coefficients of 0.9784/0.8788 and 0.9805/0.9497, respectively. MRF*, however, is only accurate for T1 measurements, while it deviates for T2, showing a T1/T2 correlation of 0.9775/0.5695. Figures 4c and 4d compares the efficiency of each method, defined as precision per square root of acquisition time, where the increase in QTI efficiency can be attributed to both factors: increased precision in reduced time. Finally, Figs. 4e and 4f show the results of parameter estimation and uncertainty quantification using $\Pi$4U for Bayesian inference[39]. Table 1 compares QTI against other 2D MRF variants, where it can be observed that QTI obtains information from marginally smaller voxels and with similar speed as the optimized version of Assländer *et al.*[14], with the added benefit of parameter uncertainty quantification and creation of angiography images. Also, while the examples shown here do not explicitly encode for transmit field inhomogeneities through B1+ mapping[28,47], our formulation allows us to incorporate separate B1+ maps prior to parameter inference. Thus, by using fast B1+ mapping methods[48], the methods presented here can be used at any field strength with increased efficiency. To ensure consistency with these previous MRF variants, no other physical effects such as B0 inhomogeneities, diffusion, or magnetisation transfer were considered in our analysis.

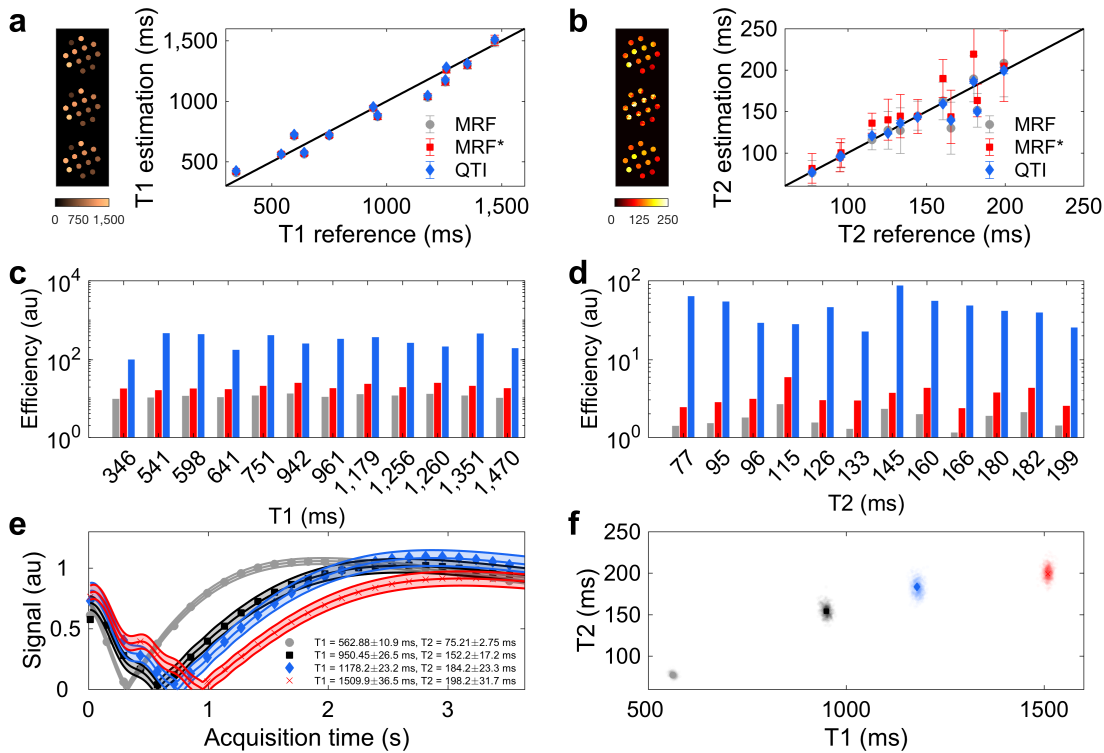

*Figure 4: Phantom validation and uncertainty quantification. a-b*, Measurement accuracy for QTI and MRF with respect to the reference. Measurements show mean ± standard deviation over 5,000 voxels for each tube obtained from regions of interest from 10 independent measurements. MRF* refers to MRF truncated to a 3.66 second acquisition and the inset on the left of each image shows, from top to bottom, the estimated parametric maps in the phantom for MRF, MRF*, and QTI. *c-d,* Efficiency of each individual method. Efficiency is determined as precision (mean / standard deviation) per the square root of the acquisition time. *e*, Transient-state signal evolution for four representative voxels in the phantom. The discrete points represent measured data points, while the continuous lines are the signals which best fit the data ± the corresponding uncertainty. Here, we can observe a trend of increased uncertainty with increasing T1/T2 values. *f*, Posterior probability density function of the four representative voxels in parameter space and maximum likelihood estimation. The posterior density function becomes broader as the parameters increase, showing again the trend of increased uncertainty with increasing parameter values. This also gives an indication of parameter encoding: the lower the uncertainty, the better the corresponding parameters are encoded into the signal. QTI is thus an accurate and efficient method for multiparametric estimation and uncertainty quantification.

**Table 1: Resolution, scan time, and capabilities of QTI versus different 2D MRF variants.**

|  | QTI | MRF | | | |
|---|---|---|---|---|---|
|  |  | Ma *et al.* [7] | Jiang *et al.* [40] | Cloos *et al.* [49] | Assländer *et al.* [14] |
| Resolution (mm$^3$) | 1.2 x 1.2 x 2 | 2.3 x 2.3 x 5 | 1.17 x 1.17 x 5 | 1.4 x 1.4 x 5 | 1 x 1 x 3 |
| Scan time per slice (s) | 3.66 | 12 | 13 | 7-21 | 3.80 |
| Clinical parameters | T1, T2, PD | T1, T2, PD | T1, T2, PD | T1, T2, PD | T1, T2, PD |
| B1$^+$ mapping | No | No | No | Yes | No |
| Parameter uncertainty | Yes | No | No | No | No |
| Angiography | Yes | No | No | No | No |

**In vivo evaluation.** We additionally validated our approach with volunteer data, where the obtained quantification proved consistent with literature[50–56] findings for different tissue types in the brain (Supplementary Table 1). Figure 5 shows the results of the reconstructed data, from which we sum the absolute value of frames to obtain angiography images (Fig. 6) and make use of the signal dynamics of the reconstructed dataset to perform probabilistic parameter inference for the simultaneous estimation of PD, T1, and T2. Also, we compared the vasculature data obtained with QTI versus a standard clinical time of flight angiography pulse sequence, where the only notable difference is that QTI's increased sensitivity to both slow and fast flowing blood in arteries and veins (Fig. 7, Supplementary Fig. 3, and Supplementary Video 1-2). This additional information could significantly impact the diagnosis of stroke and other vascular diseases. Importantly, for the patient and the radiologist these projections can be attained without additional specific angiography sequences at the expense of increased scan times: whereas conventional techniques would require serial measurements ranging from 30 minutes to one hour to obtain the same information, here the entire 4D set of complementary data is obtained in under seven minutes.

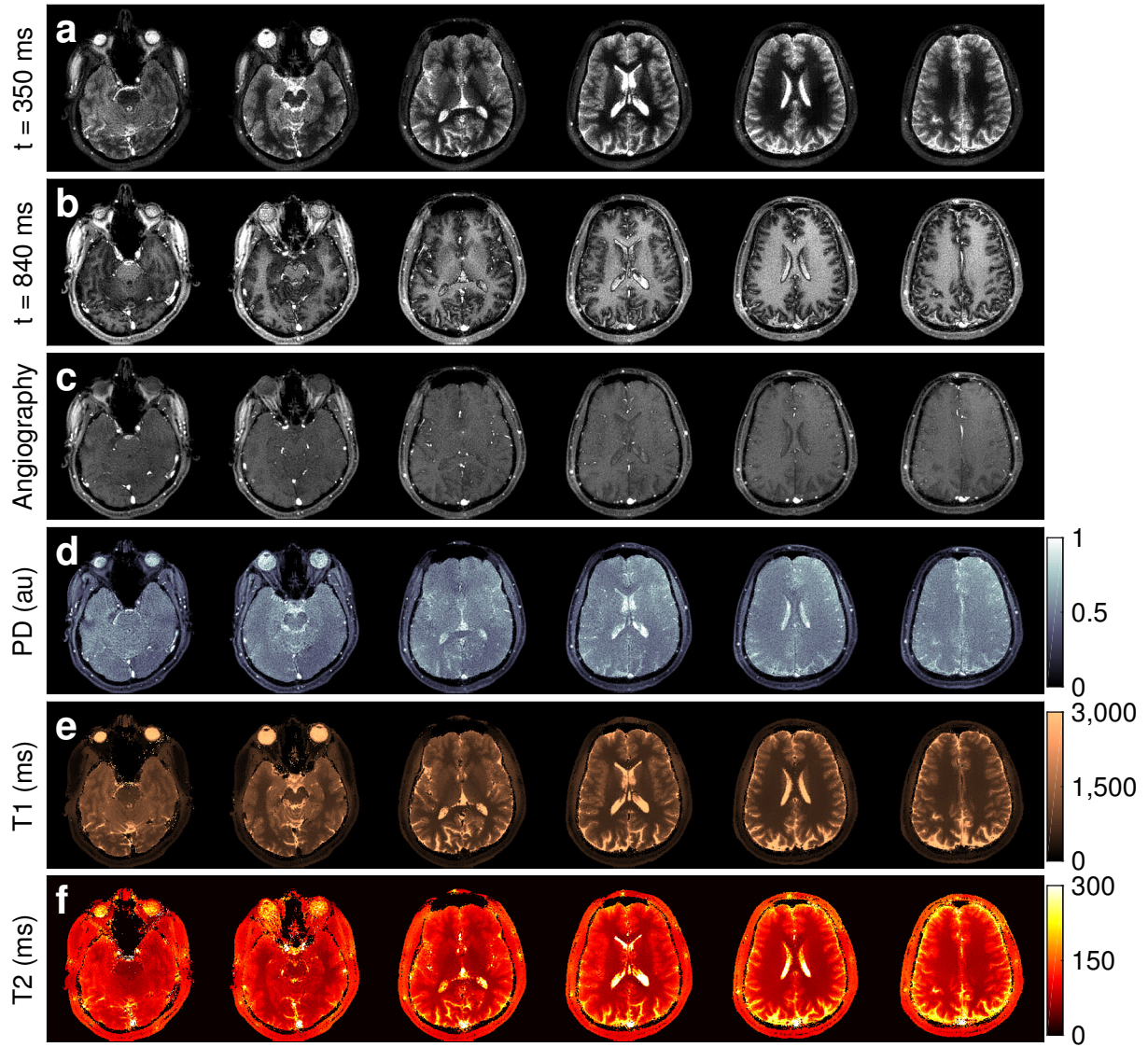

*Figure 5: Simultaneous contrast-weighted imaging, angiography, and multiparametric mapping. **a-b,** The proposed reconstruction framework allows us to recover a series of contrast-weighted images which follow the signal dynamics given by the transient-state model. For example, one can observe contrast inversion between WM and GM as these signals evolve differently in time (see also Fig. 3b). **c,** By design, combining frames of the reconstructed data creates signal hyperintensities which can be used to obtain vascular information, such as angiography projections (see Fig. 6). **d-f,** Parameter inference techniques enable the computation of the voxel-wise parametric maps that best describe the signal evolutions. The entire 4D dataset of 112, 2 mm thick slices with 1.2 mm$^2$ in-plane resolution was acquired in 6:48.*

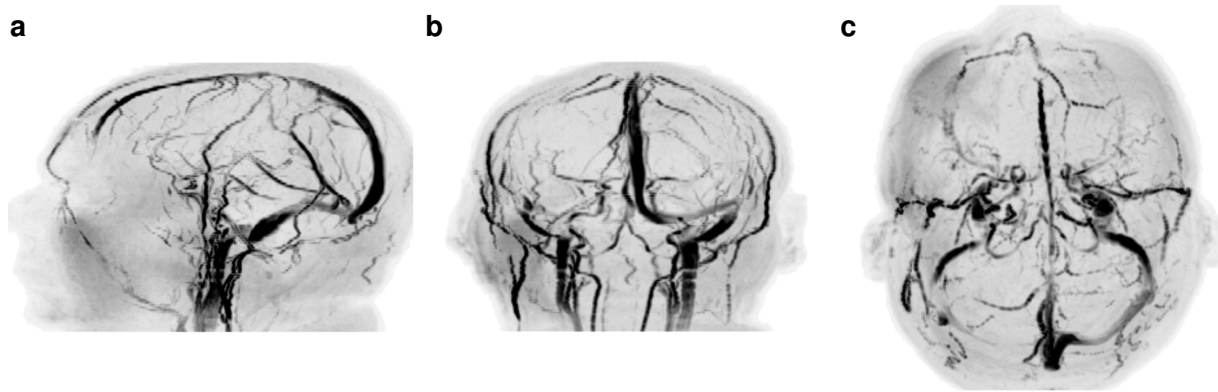

*Figure 6: Angiography projections. Maximum intensity projections obtained by combining the last 160 frames of the reconstructed data (Fig. 5c) results in vasculature images. The angiography allows for an immediate visualization of all the main vascular structures in the head, such as the Circle of Willis, the carotid arteries, and the superior sagittal sinus. The projections show a left (**a**) and sagittal view, a coronal view (**c**), and an axial view (**d**). This information is directly obtained from the reconstructed measurements without the need for separate vascular scans.*

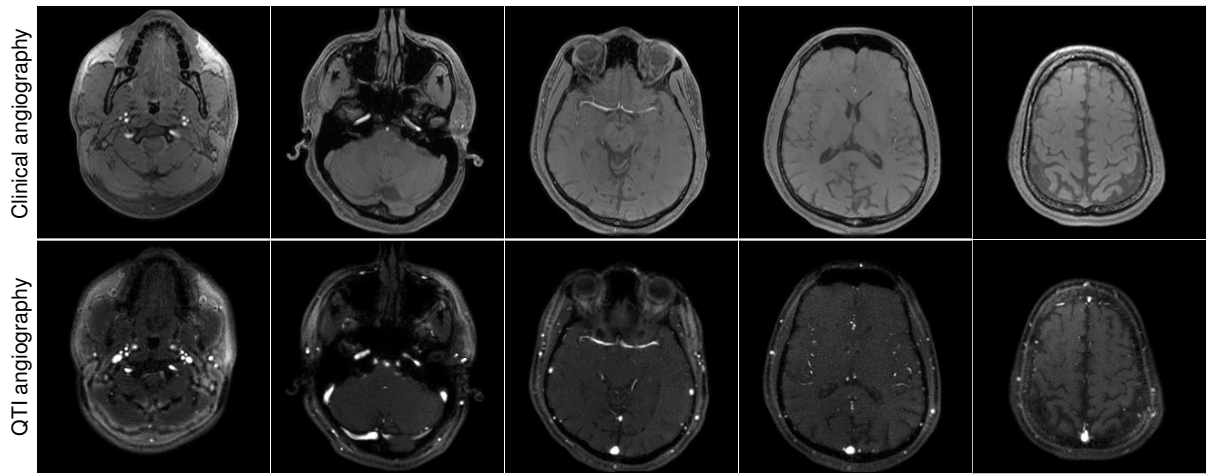

*Figure 7: Validation versus standard clinical angiography. A comparison between the vascular information obtained from the reconstructed data shows a high degree of correspondence with a standard clinical time of flight angiography sequence, with the difference that QTI has increased sensitivity to slow blood flow in both arteries and veins. A maximum intensity projection over the data shown here produces the visualisations in Fig. 6, validating our approach as a novel method for creating imaging biomarkers from vascular information.*

## Discussion

As MR sequences are developed towards multi-contrast, turnkey solutions, the methods described here offer a roadmap to systematically encode for additional parameters while simultaneously adding valuable image biomarkers derived from non-encoded parameters that affect the acquired signal. This was achieved by incorporating flow into the signal model, developing a design framework that accounts for signal contrast and parameter encoding, reconstructing space-resolved temporal signals, and performing a comprehensive analysis on these signals.

In theory, a quantification of the velocity of flowing blood is also possible with our proposed sequence. However, our sequence does not explicitly encode for velocity and the constant velocity scalar used in our model oversimplifies the complex reality of flow inside vessels, leading to parametric information with unclear clinical utility (Supplementary Fig. 4). Moreover, the model accounts only for flow perpendicular to the slice, while failing to capture movement within the slice. Incorporating flow sensitive gradients in the design process and adjusting the model to account for more complex motion and perfusion phenomena is a possibility to extract flow information inherent in the data, and this is the subject of future work. On the other hand, this model, however simple, proved useful in the context of contrast maximisation: it provided vasculature images that could be immediately used in clinical routines. The creation of these images comes with no additional penalty on acquisition times; on the contrary, this is the fastest demonstration of multiparametric mapping in the transient-state – and the only capable of additionally obtaining relevant angiography images.

Several works have considered optimising sequences, either based on Cramér-Rao lower bounds or other metrics derived from the Fisher information matrix[14,57–59], or by designing to maximise contrast for particular tissue classes[60]. Equation 5, conversely, considers both parameter encoding and contrast maximisation. Moreover, by using a prior, the optimization is not limited to a single sample of parameters or a restricted range specified from a training set. Rather, the design problem can be tackled in a multiclass, multiparametric fashion, ensuring the optimised design performs well for a wide range of values. Also, the information provided by the tissue priors can be used to incorporate additional design constraints, such as contrast maximisation of tissue classes. Thereafter, it is straightforward to extended this design framework to other uses by either modifying the prior to be e.g. specific to certain diseases, enforcing higher contrast between certain classes, or focusing exclusively on parameter encoding.

Both our Bayesian design framework and Bayesian parameter inference techniques operate under the simplifying assumption of a Gaussian noise model. Although these assumptions may be violated by combining complex data from multiple receiver coils and reconstructing using low rank approximations, we observed only a small model error with 10 subspace coefficients (Supplementary Fig. 5) and no significant differences when comparing against other inference techniques, such as dictionary matching or simple least squares fitting (Supplementary Fig. 6). Moreover, the uncertainty provided by Bayesian inference resulted in additional information regarding sequences encoding and parameter quantification. Also, the subspace images on themselves offer discriminative information between different classes (Supplementary Fig. 7), and could be used for subsequent image processing tasks, such as tissue segmentation.

## Conclusion

This comprehensive information offered by QTI could provide valuable insight in subsequent precision medicine applications, as multi-contrast, multiparametric data improves f.eg. the automated classification of the aggressiveness of prostate cancer[61], the metastatic potential of melanomas[62], or the screening for chronic diseases[63]. Moreover, our approach can be adapted to additional clinical biomarkers, where their direct measurement in parallel to the quantification of multiple parametric maps offers novel sources of information, opening new dimensions that will facilitate quantitative in-vivo phenotyping, improving diagnostic capabilities and facilitating the study and understanding of the human body in health and disease.

## Acknowledgements

With the support of the TUM Institute for Advanced Study, funded by the German Excellence Initiative and the European Commission under Grant Agreement Number 605162 and 656937. We would like to thank Georgios Arampatzis for his support and feedback on $\Pi4U$.

## Author contributions

P.A.G., M.M., G.B., M.I.M. and B.H.M. developed the concept and designed the research; P.A.G., M.M., and G.B. performed the technical implementation; P.A.G. and M.M. collected and analysed the data; all authors contributed to manuscript editing and development.

## Competing Interests

M.I.M. is an employee at GE Healthcare, Munich, Germany. All other authors declare no competing interests.

# References

1. Lauterbur, P. C. Image formation by induced local interactions. Examples employing nuclear magnetic resonance. *Nature* **242,** 190–191 (1973).
2. Tofts, P. *Quantitative MRI of the Brain: Measuring Changes Caused by Disease*. *Wiley* (Wiley, 2003).
3. Fisher, M. New Magnetic Resonance Techniques for Acute Ischemic Stroke. *JAMA J. Am. Med. Assoc.* **274,** 908 (1995).
4. Polak, J. F. *et al.* Detection of internal carotid artery stenosis: comparison of MR angiography, color Doppler sonography, and arteriography. *Radiology* **182,** 35–40 (1992).
5. Stankovic, Z., Allen, B. D., Garcia, J., Jarvis, K. B. & Markl, M. 4D flow imaging with MRI. *Cardiovasc. Diagn. Ther.* **4,** 173–92 (2014).
6. Schnell, S., Wu, C. & Ansari, S. A. Four-dimensional MRI flow examinations in cerebral and extracerebral vessels - ready for clinical routine? *Curr. Opin. Neurol.* **29,** 419–28 (2016).
7. Ma, D. *et al.* Magnetic resonance fingerprinting. *Nature* **495,** 187–192 (2013).
8. Warntjes, J. B. M., Dahlqvist, O. & Lundberg, P. Novel method for rapid, simultaneous T1, T2*, and proton density quantification. *Magn. Reson. Med.* **57,** 528–537 (2007).
9. Gras, V., Farrher, E., Grinberg, F. & Shah, N. J. Diffusion-weighted DESS protocol optimization for simultaneous mapping of the mean diffusivity, proton density and relaxation times at 3 Tesla. *Magn. Reson. Med.* **78,** 130–141 (2017).
10. Deoni, S. C. L., Peters, T. M. & Rutt, B. K. High-resolution T1 and T2 mapping of the brain in a clinically acceptable time with DESPOT1 and DESPOT2. *Magn. Reson. Med.* **53,** 237–241 (2005).
11. Ehses, P. *et al.* IR TrueFISP with a golden-ratio-based radial readout: Fast quantification of T1, T2, and proton density. *Magn. Reson. Med.* **69,** 71–81 (2013).
12. Schmitt, P. *et al.* Inversion recovery TrueFISP: quantification of T(1), T(2), and spin density. *Magn. Reson. Med.* **51,** 661–667 (2004).
13. Stöcker, T. *et al.* MR parameter quantification with magnetization-prepared double echo steady-state (MP-DESS). *Magn. Reson. Med.* **72,** 103–111 (2014).
14. Assländer, J., Lattanzi, R., Sodickson, D. K. & Cloos, M. A. Relaxation in Spherical Coordinates: Analysis and Optimization of pseudo-SSFP based MR-Fingerprinting. *arXiv eprint* (2017).
15. Christen, T. *et al.* MR vascular fingerprinting: A new approach to compute cerebral blood volume, mean vessel radius, and oxygenation maps in the human brain. *Neuroimage* **89,** 262–70 (2014).
16. Anderson, C. E. *et al.* Dual Contrast - Magnetic Resonance Fingerprinting (DC-MRF): A Platform for Simultaneous Quantification of Multiple MRI Contrast Agents. *Sci. Rep.* **7,** 8431 (2017).
17. Xiaozhi Cao *et al.* A model-based velocity mapping of blood flow using MR fingerprinting. *Proc Intl Soc Mag Reson Med* (2017).
18. Flassbeck, S. *et al.* Quantification of Flow by Magnetic Resonance Fingerprinting. *Proc Intl Soc Mag Reson Med* (2017).
19. Weigel, M. Extended phase graphs: Dephasing, RF pulses, and echoes - pure and simple. *J. Magn. Reson. Imaging* (2014). doi:10.1002/jmri.24619
20. Hennig, J. Echoes—how to generate, recognize, use or avoid them in MR-imaging sequences. Part I: Fundamental and not so fundamental properties of spin echoes. *Concepts Magn. Reson.* **3,** 125–143 (1991).
21. Axel, L. Blood flow effects in magnetic resonance imaging. *Am.J.Roentgeno.* **143,** 1157–1166 (1984).

22. Assländer, J. *et al.* Low Rank Alternating Direction Method of Multipliers Reconstruction for MR Fingerprinting. *arXiv eprint* arXiv:1608.06974 (2016). doi:10.1002/mrm.26639
23. Hamilton, J. I. *et al.* Pulse Sequence Optimization for Improved MRF Scan Efficiency. *Proc Intl Soc Mag Reson Med* **23,** 3386 (2015).
24. Cohen, O. Magnetic Resonance Fingerprinting Trajectory Optimization. *Proc Intl Soc Mag Reson Med* **1,** 1 (2014).
25. Cohen, O., Sarracanie, M., Rosen, M. S. & Ackerman, J. L. In Vivo Optimized Fast MR Fingerprinting in the Human Brain. *Proc Intl Soc Mag Reson Med* (2016).
26. Verdinelli, I. & Chaloner, K. Bayesian Experimental Design : A Review. *Stat. Sci.* **10,** 273–304 (1995).
27. Owen, D. *et al.* Optimisation of Arterial Spin Labelling Using Bayesian Experimental Design. *MICCAI Int. Conf. Med. Image Comput. Comput. Interv.* **9902,** 511–518 (2016).
28. Buonincontri, G. & Sawiak, S. Three-dimensional MR fingerprinting with simultaneous B1 estimation. *Magn. Reson. Med.* **00,** 1–9 (2015).
29. Ulas, C. *et al.* Robust Reconstruction of Accelerated Perfusion MRI Using Local and Nonlocal Constraints. *Int. Work. Reconstr. Anal. Mov. Body Organs* **LNCS 10129,** 37–47 (2017).
30. Tamir, J. I. *et al.* T2 shuffling: Sharp, multicontrast, volumetric fast spin-echo imaging. *Magn. Reson. Med.* (2016). doi:10.1002/mrm.26102
31. Lingala, S. G. & Jacob, M. Blind Compressive Sensing Dynamic MRI. *IEEE Trans. Med. Imaging* **32,** (2013).
32. Ulas, C., Gomez, P. A., Sperl, J. I., Preibisch, C. & Menze, B. H. Spatio-temporal MRI Reconstruction by Enforcing Local and Global Regularity via Dynamic Total Variation and Nuclear Norm Minimization. *Proc. Int. Symp. Biomed. Imaging* 306–309 (2016). doi:10.1109/ISBI.2016.7493270
33. Davies, M., Puy, G., Vandergheynst, P. & Wiaux, Y. A Compressed Sensing Framework for Magnetic Resonance Fingerprinting. *SIAM J. Imaging Sci.* **7,** 2623–2656 (2014).
34. Zhao, B. *et al.* A Model-Based Approach to Accelerated Magnetic Resonance Fingerprinting Time Series Reconstruction. *Proc Intl Soc Mag Reson Med* (2016).
35. Gómez, P. A. *et al.* Learning a Spatiotemporal Dictionary for Magnetic Resonance Fingerprinting with Compressed Sensing. *MICCAI Patch-MI Work.* **LNCS 9467,** 112–119 (2015).
36. Hargreaves, B. A., Nishimura, D. G. & Conolly, S. M. Time-optimal multidimensional gradient waveform design for rapid imaging. *Magn. Reson. Med.* **51,** 81–92 (2004).
37. Fessler, J. A. & Sutton, B. P. Nonuniform Fast Fourier Transforms Using Min-Max Interpolation. *IEEE Trans Signal Process.* **51,** 560–574 (2003).
38. Walsh, D. O., Gmitro, A. F. & Marcellin, M. W. Adaptive reconstruction of phased array MR imagery. *Magn. Reson. Med.* **43,** 682–90 (2000).
39. Hadjidoukas, P. E., Angelikopoulos, P., Papadimitriou, C. & Koumoutsakos, P. Π4U: A high performance computing framework for Bayesian uncertainty quantification of complex models. *J. Comput. Phys.* **284,** 1–21 (2015).
40. Jiang, Y., Ma, D., Seiberlich, N., Gulani, V. & Griswold, M. A. MR Fingerprinting Using Fast Imaging with Steady State Precession (FISP) with Spiral Readout. *MRM* (2014).
41. Gómez, P. A. *et al.* 3D Magnetic Resonance Fingerprinting with a Clustered Spatiotemporal Dictionary. *Proc Intl Soc Mag Reson Med* (2016).
42. Hamilton, J. I., Deshmane, A., Hougen, S., Griswold, M. A. & Seiberlich, N. Magnetic Resonance Fingerprinting with Chemical Exchange (MRF-X) for Quantification of Subvoxel T1, T2, Volume Fraction, and Exchange Rate. *Proc Intl Soc Mag Reson Med* **23,** 16728 (2015).
43. Buonincontri, G., Schulte, R., Cosottini, M., Sawiak, S. & Tosetti, M. Spiral MRF at 7T with simultaneous B1 estimation. *Proc Intl Soc Mag Reson Med* (2016).

44. Jiang, Y. *et al.* Repeatability of magnetic resonance fingerprinting T1 and T2 estimates assessed using the ISMRM/NIST MRI system phantom. *Magn. Reson. Med.* (2016). doi:10.1002/mrm.26509
45. Lustig, M., Donoho, D. & Pauly, J. M. Sparse MRI: The application of compressed sensing for rapid MR imaging. *Magn. Reson. Med.* **58,** 1182–1195 (2007).
46. Lustig, M., Donoho, D. L., Santos, J. M. & Pauly, J. M. Compressed Sensing MRI. *IEEE Signal Process. Mag.* **25,** (2008).
47. Cloos, M. A. *et al.* Multiparametric imaging with heterogeneous radiofrequency field: Additional material. *Nat. Commun.* **53,** 1689–1699 (2016).
48. Nehrke, K. & Börnert, P. DREAM-a novel approach for robust, ultrafast, multislice B1 mapping. *Magn. Reson. Med.* **68,** 1517–1526 (2012).
49. Cloos, M. A. *et al.* Multiparamatric imaging with heterogenous radiofrequency fields. *Nat. Commun.* 1–10 (2016). doi:10.1038/ncomms12445
50. Lu, H. *et al.* Routine clinical brain MRI sequences for use at 3.0 tesla. *J. Magn. Reson. Imaging* **22,** 13–22 (2005).
51. Ethofer, T. *et al.* Comparison of Longitudinal Metabolite Relaxation Times in Different Regions of the Human Brain at 1.5 and 3 Tesla. *Magn. Reson. Med.* **50,** 1296–1301 (2003).
52. Gelman, N. *et al.* MR Imaging of Human Brain at 3.0 T: Preliminary Report on Transverse Relaxation Rates and Relation to Estimated Iron Content. *Radiology* **210,** 759–767 (1999).
53. MacKay, A. *et al.* Insights into brain microstructure from the T2 distribution. *Magn. Reson. Imaging* **24,** 515–525 (2006).
54. Smith, S. A., Edden, R. A. E., Farrell, J. A. D., Barker, P. B. & Van Zijl, P. C. M. Measurement of T1 and T2 in the cervical spinal cord at 3 Tesla. *Magn. Reson. Med.* **60,** 213–219 (2008).
55. Whittall, K. P. *et al.* In vivo measurement ofT2 distributions and water contents in normal human brain. *Magn. Reson. Med.* **37,** 34–43 (1997).
56. Noeske, R., Seifert, F., Rhein, K. H. & Rinneberg, H. Human cardiac imaging at 3 T using phased array coils. *Magn. Reson. Med.* **44,** 978–82 (2000).
57. Zhao, B., Haldar, J. P., Setsompop, K. & Wald, L. L. Optimal experiment design for magnetic resonance fingerprinting. in *2016 38th Annual International Conference of the IEEE Engineering in Medicine and Biology Society (EMBC)* 453–456 (IEEE, 2016). doi:10.1109/EMBC.2016.7590737
58. Teixeira, R. P. A. G., Malik, S. J. & Hajnal, J. V. Joint system relaxometry (JSR) and Crámer-Rao lower bound optimization of sequence parameters: A framework for enhanced precision of DESPOT $T_1$ and $T_2$ estimation. *Magn. Reson. Med.* **79,** 234–245 (2018).
59. Rochefort, R. & De, valabregue and L. Fisher Information Matrix for Optimizing the Acquisition Parameters in Multi-Parametric Mapping Based on Fast Steady-State Sequences. *Proc Intl Soc Mag Reson Med* (2016).
60. Van Reeth, E. *et al.* Optimal control design of preparation pulses for contrast optimization in MRI. *J. Magn. Reson.* **279,** 39–50 (2017).
61. Fehr, D. *et al.* Automatic classification of prostate cancer Gleason scores from multiparametric magnetic resonance images. *Proc. Natl. Acad. Sci. U. S. A.* **112,** E6265-73 (2015).
62. Li, L. Z. *et al.* Quantitative magnetic resonance and optical imaging biomarkers of melanoma metastatic potential. *Proc. Natl. Acad. Sci. U. S. A.* **106,** 6608–13 (2009).
63. Perkins, B. A. *et al.* Precision medicine screening using whole-genome sequencing and advanced imaging to identify disease risk in adults. *Proc. Natl. Acad. Sci. U. S. A.* **115,** 3686–3691 (2018).